# PtIr protective coating system for precision glass molding tools: design, evaluation and mechanism of degradation

Marcel Friedrichs[1], Zirong Peng[2,]*, Tim Grunwald[1], Michael Rohwerder[2], Baptiste Gault[2,3] and Thomas Bergs[1]

[1]Fraunhofer Institute for Production Technology IPT, Steinbachstraße 17, 52074 Aachen, Germany
[2]Max-Planck-Institut für Eisenforschung GmbH, Max-Planck-Straße 1, 40237 Düsseldorf, Germany
[3]Department of Materials, Royal School of Mines, Imperial College, London, SW7 2AZ, UK

*corresponding author: z.peng@mpie.de

**Keywords:** metal coatings, physical vapor deposition, degradation mechanism, diffusion, oxidation, Precision Glass Molding

**Abstract**

During Precision Glass Molding (PGM), the molding tools have to withstand severe thermo-chemical and thermo-mechanical loads cyclically. To protect their high-quality optical surface against degradation and increase their service lifetime, protective coatings are applied on the molding tools. In this work, we designed four different PtIr protective coating systems, where the thickness of the PtIr layer and the adhesion layer were varied. Their lifetimes were evaluated and compared using an in-house built testing bench. Among all the studied coating systems, the protective coating, which consists of a 600-nm-thick PtIr layer and a 20-nm-thick Cr adhesion layer, showed the best durability with the longest lifetime. To understand the degradation mechanism of the coating during actual engineering production, an industrial PGM machine was used and emulation PGM tests were conducted. Detailed sample characterization was performed using an array of complementary techniques including white light interferometry (WLI), scanning electron microscopy (SEM), energy-dispersive X-ray spectroscopy (EDX), scanning transmission electron microscopy (STEM) and atom probe tomography (APT).

Phenomena such as interdiffusion, oxidation, coating spallation and glass sticking on the coating were observed and are discussed in the context of optimization of the coating's performance and durability.

**1. Introduction**

High-quality glass lenses with complex geometries are at the core of innovative products in a wide range of applications such as laser technology, medical technology and consumer electronics [1–4]. For production of such lenses, precision glass molding (PGM) is an established and economical manufacturing technology [5–7]. In contrast to direct grinding and polishing processes, PGM is a replicative technology. PGM is suitable for medium and high volume production of complex glass components as aspherical lenses (i.e. aspheres) or freeform lenses. During PGM, a glass blank is heated above the glass-transition temperature into the viscous state, and then the blank is molded between two high-precision molding tools. Upon controlled cooling, the molded glass lens retains the desired shape along with a high surface quality without any post-processing steps [8].

Commonly, the molding tools are made of cemented tungsten carbide, one of the most frequently applied tool materials in industry [9]. Combining the hard and brittle WC grains with soft and ductile metallic binders, cemented tungsten carbides exhibit high strength, hardness and wear resistance even at elevated temperatures [10]. Furthermore, WC can be ground and polished to a surface quality required for molding high-precision glass components [11]. To protect the tungsten carbide main part from damages induced by the harsh operation conditions, a surface protective coating is usually employed [12]. The surface must withstand not only thermal and mechanical stresses, but also corrosive thermo-chemical environment due to the direct contact with the hot aggressive molten glass. The key requirements for such protective coating are: (1) good thermal stability, i.e. no aging at molding temperatures (350–800 °C), (2) strong and hard enough, with sufficient resistance to thermal shocks, (3) chemically highly stable and inactive towards reactions with glass.

Three different groups of surface protective coatings are available for this application, including (1) metal or precious metal alloy coatings such as Pt-Ir [13], Ir-Re [14] and Mo-Ru [15] alloys, (2) ceramic coatings such as TiN [16], TiAlN [17] and CrN [18], and (3) carbon-based coatings such as diamond-like carbon (DLC) [19]. Among them, precious metal alloy coatings with superior oxidation and corrosion resistance are suggested to be most reliable and are currently widely used industrially [20,21]. Therefore, in this work, we focused our attention on the Pt-Ir coating with approx. 30 at.% Pt and 70 at.% Ir, referred to as PtIr across this article. This composition has been selected for a compromise between the mechanical and chemical properties. With the increase in the Pt content, its oxidation resistance also increases, however the hardness of the film significantly decreases [22]. To enhance the adhesion strength of the PtIr coatings on the cemented tungsten carbide main part, an adhesion layer is also applied, which is normally a pure metallic thin film. Most frequently used ones are Cr and Ni [12,17,23]. We chose a Cr adhesion layer here because it was found to exhibit better performance than Ni [24–26]. To find the optimal thickness of the PtIr and Cr adhesion layer, four different coating systems, namely a 300-nm-thick PtIr layer with a 20-nm-thick Cr layer (referred to as 300PtIr/20Cr), a 600-nm-thick PtIr layer with a 20-nm-thick Cr layer (referred to as 600PtIr/20Cr), a 600-nm-thick PtIr layer with a 5-nm-thick Cr layer (referred to as 600PtIr/5Cr), and a 600-nm-thick PtIr without Cr layer (referred to as 600PtIr), were designed. An industrial PGM machine (GMP-211V,Toshiba Machine Co., Ltd.,) as well as a home-built lifetime testing bench [27] were employed to examine the coating systems.

To reveal the detailed degradation mechanism of the coating system, a variety of complementary characterization techniques were applied, including white light interferometry (WLI) for surface morphology and roughness analysis, and scanning electron microscopy (SEM) combined with energy-dispersive X-ray spectroscopy (EDX) for surface defects analysis. Furthermore, we employed atom probe tomography (APT) and scanning transmission electron microscopy (STEM), to provide a three-dimensional distribution of elements at the atomic scale and study the microstructural evolution, so as to enable a better understanding of the internal mass transport process [25,28].

## 2. Materials and methods

### 2.1. Sample preparation

In order to study the same mold material as the ones used in PGM industrial production, we also chose commercial cemented tungsten carbide as substrate (CTN01L, Ceratizit S.A., cemented tungsten carbide with around 0.3 wt.% Co binder). As shown in Figure 1, we employed two types of samples. (a) Flat samples with an even surface were used in the industrial PGM testing process and (b) pin samples with a spherical surface were used in the lifetime testing bench. To ensure a good coating quality, cemented WC substrates were polished until their surface roughness Ra was smaller than 10 nm. Afterwards, they were cleaned in an ultrasonic bath at four different frequencies (40, 58, 75, and 132 kHz; 3 min each frequency). A CemeCon CC800/9 Custom magnetron sputtering unit was used for coating deposition at Fraunhofer IPT. Four different PtIr coating systems with various coating thicknesses and adhesion layers (Cr, no interlayer) were deposited, as illustrated in Figure 2. Details about the deposition process were reported in another publication [25].

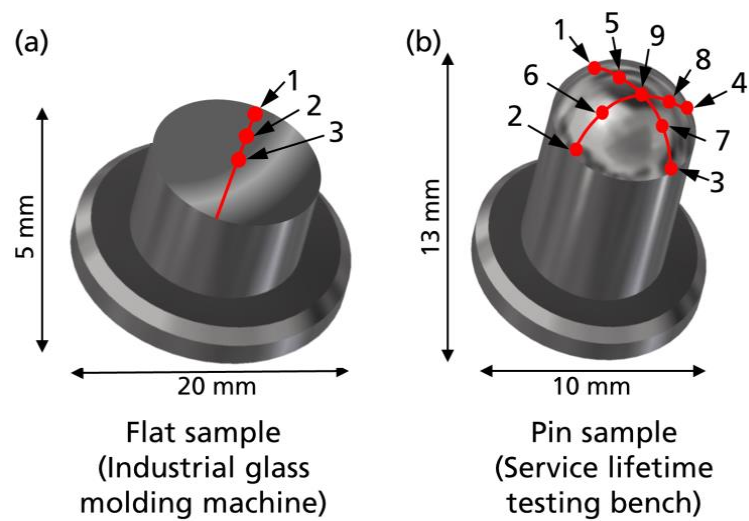

*Figure 1: Sketch to illustrate the geometry of (a) the flat sample and (b) the pin sample. The red dots indicate the positions of the white light interferometer measurements.*

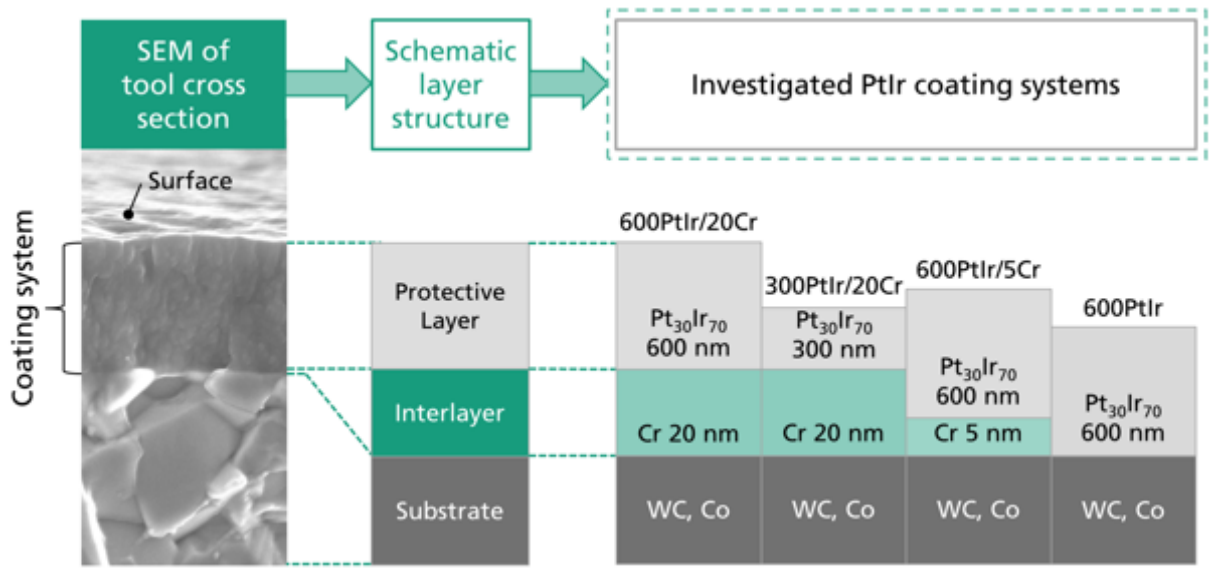

*Figure 2: Schematic illustration of the protective coating systems studied in this work.*

**2.2. Industrial precision glass molding test**

An industrial glass molding machine (GMP-211V manufactured by Toshiba Machine Co., Ltd., Japan) was employed to investigate the degradation of the coating during the actual PGM production. Figure 3 illustrates the testing setup and detailed conditions of a complete cycle of the PGM process. We used the B270® type glass (Schott AG) containing $SiO_2$, $Na_2O$, $CaO$, $K_2O$ and $Sb_2O_3$ etc. as the glass blanks. Initially, they were in a cylindrical form with a diameter of 10 mm and a height of 5 mm. After molding, their heights were reduced to 2.5 mm and their diameter were around 13 mm.

A molding cycle consists in four stages, namely heating, homogenization, molding and cooling. After inserting the glass blank, the molding chamber is evacuated and reaches a pressure below $3\times10^{-5}$ bar to remove oxygen and water vapor. The low pressure helps to mitigate subsequent oxidation of the molding tools at elevated temperatures. Then, infrared lamps are used to heat both the molding tools and glass up to the molding temperature. For B270®, the typical molding temperature adopted by industrial producers is approx. 640 °C, therefore, we selected this temperature for our experiments. The molding process is an isothermal process, which means that the molds and the glass blank are held at the same temperature during molding. Since we can only monitor the temperature of the molding tools, to ensure that the glass blank is also thoroughly heated up, i.e. the temperature distribution within it is homogeneous, the molding tool and the glass blank are held still for about 3 min before molding. This is the so-called homogenization stage. To reshape the glass blank, the lower molding tool is moved up towards the upper molding tool until the molding force (1 kN) is reached. This molding force is applied for 2 min. Subsequently, the pressing force is reduced to 0.5 kN and the system is cooled down to 520 °C, below the glass-transition temperature (533 °C), by $N_2$ gas with a relatively slow flow rate. This controlled cooling stage can improve the form accuracy of the optical product. Finally, the lower molding tool is moved downward and the whole chamber is cooled down to room temperature with a higher $N_2$ gas flow rate. The entire molding cycle takes about 23 min. After each cycle, the molded glass is manually removed and a new glass blank is inserted for the next molding cycle.

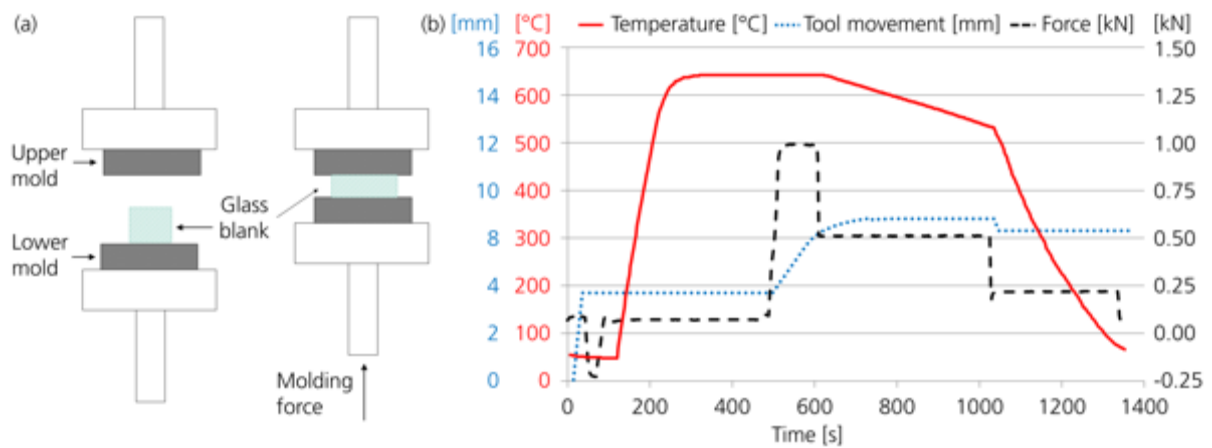

*Figure 3: (a) Setup and (b) process diagram of a cycle of the industrial precision glass molding test.*

## 2.3. Accelerated lifetime test

The lifetime of different coating systems was evaluated using a bench developed at Fraunhofer IPT [27]. Compared to an industrial glass molding machine, this facility enables us to test coatings' lifetime more efficiently and economically. On the industrial PGM machine, we can only test one molding tool pair at one time and for each PGM cycle. Reloading a new glass blank and starting the next molding cycle are performed manually. However, on the lifetime testing facility, as Figure 4 (a) shows, three molding tool samples can be examined simultaneously. They are attached to the bottom side of a pressing axis. The raw glass blanks are also of the B270® type, but in a plate shape. The length, width and thickness of the plate are 75, 45 and 6 mm respectively. 5 molding cycles can be performed with a single glass plate. For each testing round, six glass plates are used, i.e. each sample has been tested for 30 molding cycles.

The entire setup is installed in a furnace. Before testing, the atmospheric pressure of the furnace is evacuated to below $5 \times 10^{-3}$ bar. Figure 4 (b) shows the detailed process conditions of three successive molding cycles. The molding temperature is the same as the one adopted in the industrial PGM testing, i.e. 640 °C. After heating the entire setup up to the molding temperature, the mold samples are pressed into the glass plate and the molding force of 4 kN is exerted for 1 min. Then, the mold samples that are still in contact with the glass plate are cooled by a $N_2$ gas flow to 485 °C, lower than the glass-transition temperature (533 °C). Afterwards, the pressing axis with the three mold samples is elevated and one molding cycle is finished. Subsequently, the glass plates are moved automatically to the next position and a new molding cycle is started directly after a short heating period. In contrast to the industrial PGM process, here, the whole process operates in a temperature range between 640 °C and 485 °C. Upon completion of the 30 molding cycles, the entire setup is cooled down to room temperature and new glass plates are inserted for the next testing round. As mentioned previously, on the industrial PGM machine, one molding cycle takes approx. 23 min, while on the lifetime testing facility, one molding cycle is only approx. 14 min.

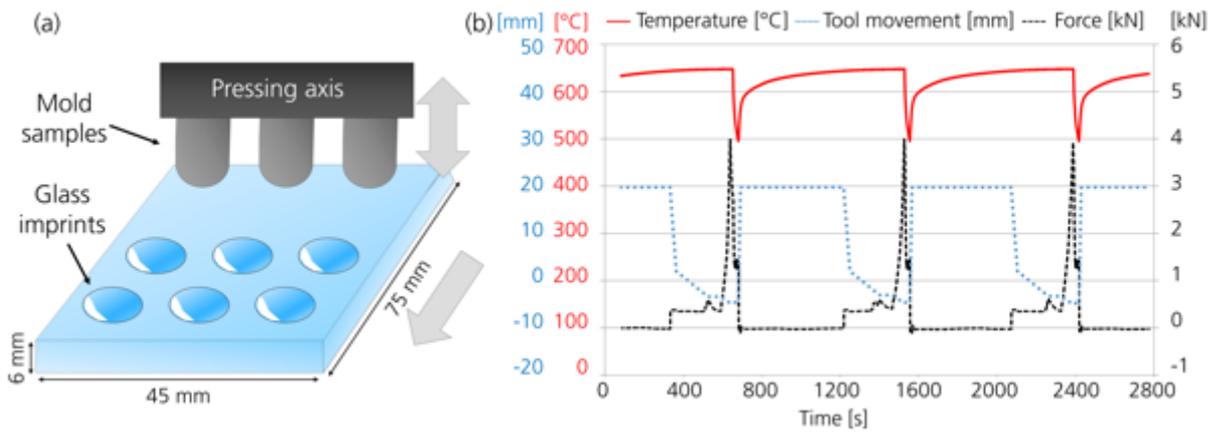

*Figure 4: (a) Setup and (b) process diagram of 3 molding cycles of the lifetime testing bench.*

**2.4. Specimen Characterization**

Firstly, light microscopy was used to access the coating quality at the macroscopic scale. Afterwards, the surface roughness Ra of the sample was measured by a white light interferometer (Contour GT-K produced by BRUKER®). During molding, the surface profile of the mold is replicated on the final molded lens. To ensure the final molded lens has a high-quality optical surface to transmit light efficiently, the surface roughness of a qualified mold must below a certain limit, which is normally 10 nm.

As Figure 1 illustrates, for the pin samples from the lifetime tests, the evaluation of Ra was based on the average value of nine measurements: four on the rim (positions 1–4), four at intermediate positions between the rim and the center (position 5–8) and, finally, one at the center (position 9). For the flat samples tested in the industrial glass molding machine, the measurements were made in a row: one on the edge (position 1), one in the middle circle (position 2) and one in the center (position 3). All the roughness measurements were made with 20 times magnification (10x2), backscan of 30 µm and length of 30 µm, VXI mode and Gaussian filter. The surface defects of the coatings were examined using SEM and EDX measurements with Zeiss Neon 40 EsB at an accelerating voltage of 10 kV.

To reveal the detailed degradation mechanism, samples from the industrial PGM process were analyzed using APT and STEM. Both APT and STEM specimens were prepared by an in-situ lift-out method in a dual beam SEM/focused ion beam (FIB) instrument (FEI Helios Nanolab 600i) [14,15]. Before the lift-out, a thin layer of Pt was deposited above the target region using electron beam assisted site-specific chemical vapor deposition, in order to protect the surface from damage by the energetic ions of the FIB. APT measurements were performed on a Cameca Instruments LEAP™ 3000X HR in laser pulsing mode with 0.8 nJ laser pulse energy and 250 kHz pulse repetition rate. The specimen base temperature was kept at 60 K and the target detection rate was set at 2 detection events / 1000 pulses. These APT experimental parameters were optimized according to a previous study [16]. STEM observations were performed using a JEOL JEM-2200FS microscope with 200 kV accelerating voltage.

## 3. Results and Discussion

### 3.1. Degradation mechanism at the early degradation stage during industrial precision glass molding test

Figure 5 shows several photos (a) and surface roughness measurements (b) of a flat mold sample with the 600PtIr/20Cr coating after an increasing number of cycles of the industrial PGM test. Since the diameter of the mold sample, the glass blank before and after the PGM process are 20, 10 and 13 mm respectively, the center region of the mold, marked as 3 in Figure 1 (a), was in contact with the glass during the whole molding cycle, while the edge area of the mold, marked as 1 in Figure 1 (a), has no contact with the glass at all. The middle position, marked as 2 in Figure 1 (a), was only in contact with the glass during the molding and cooling stages. As Figure 5 (b) illustrates, with the increase in molding cycle, the surface roughness of the mold sample increased gradually. After 120 cycles, the surface roughness of the mold sample reached around 7.5 nm, which is close to the upper limit of the surface roughness requirement for a qualified mold (10 nm). Although there was no distinct difference between the roughness values of these three different regions, a clear color change can be observed, which is an indication of surface degradation. To reveal the degradation mechanism of the early stage, we closely analyzed this 120-PGM-cycle-tested mold samples as well as the as-fabricated sample using high-resolution characterization techniques including APT and STEM.

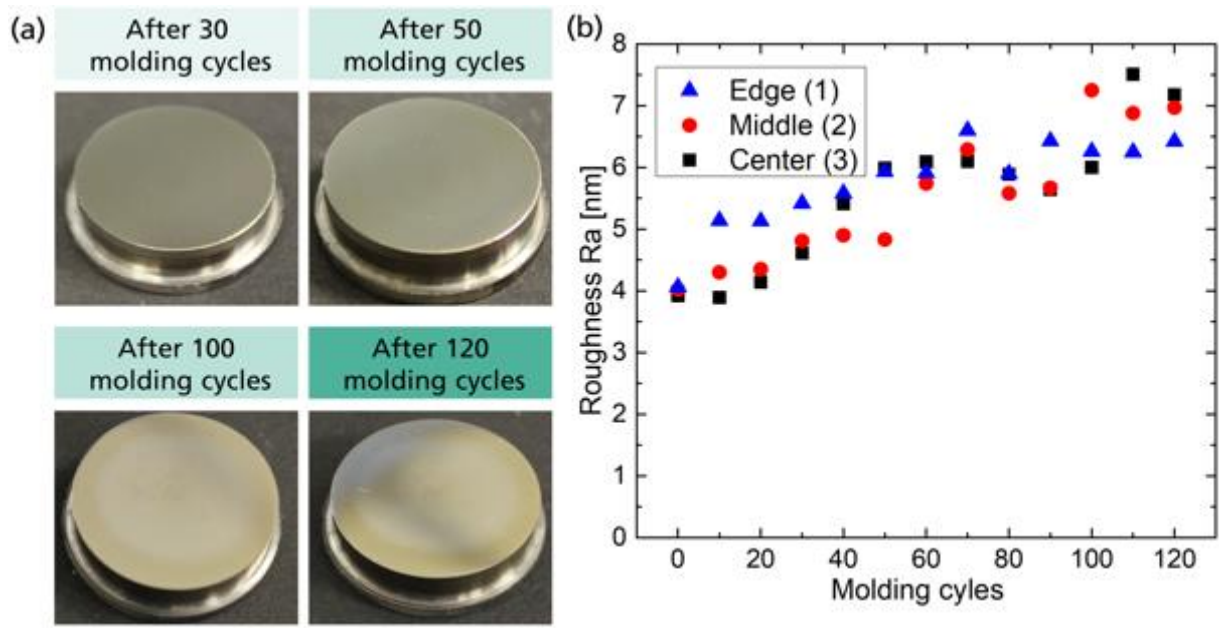

*Figure 5: (a) Photos and (b) surface roughness of a flat mold sample with the 600PtIr/20Cr coating after various molding cycles.*

Figure 6 (a) is a cross-sectional bright field STEM image of the as-fabricated 600PtIr/20Cr mold sample. The PtIr, Cr layer as well as the WC substrate can be clearly distinguished. The interfaces between them are sharp and flat. The PtIr layer is dense and void-free. It exhibits a fine columnar microstructure, which is typical for sputter-deposited thin layers. The widths of the columnar grains range from 50-100 nm. Figure 6 (b) shows an APT map with the corresponding 1D composition-depth profile obtained from the surface region of the sample, from which we can confirm that the surface of the as-fabricated sample is without any oxides or contamination. The PtIr layer has very high purity. As we explained in a previous publication [25], the in-house designed PtIr target is chemically inhomogeneous, therefore, the deposited PtIr layer also show a non-uniform composition. The Pt and Ir contents vary in the range of 26 to 36 and 63 to 73 at.% respectively.

Figure 6 (c) shows an APT map with the corresponding 1D composition-depth profile obtained near the Cr layer. Conversely to the PtIr layer, the Cr layer is not pure and contains some nanosized dispersed Cr oxides, as marked using the 6 at.% CrO iso-composition surfaces (in black). These Cr oxides particles are considered to result from the oxidation of Cr during the deposition process, as the affinity of Cr for oxygen is high. The 1D composition-depth profile reveals that at both the PtIr/Cr and Cr/WC interfaces, the compositions of Cr change rapidly, i.e. without noticeable interfacial segregation. To better illustrate the distribution of different elements within the sample, 2D jet color composition maps of Ir, Cr, and W are also plotted in Figure 6 (c), where blue indicates the lowest content and red indicates the highest content. We can confirm that there is no interdiffusion between different regions at the as-fabricated state.

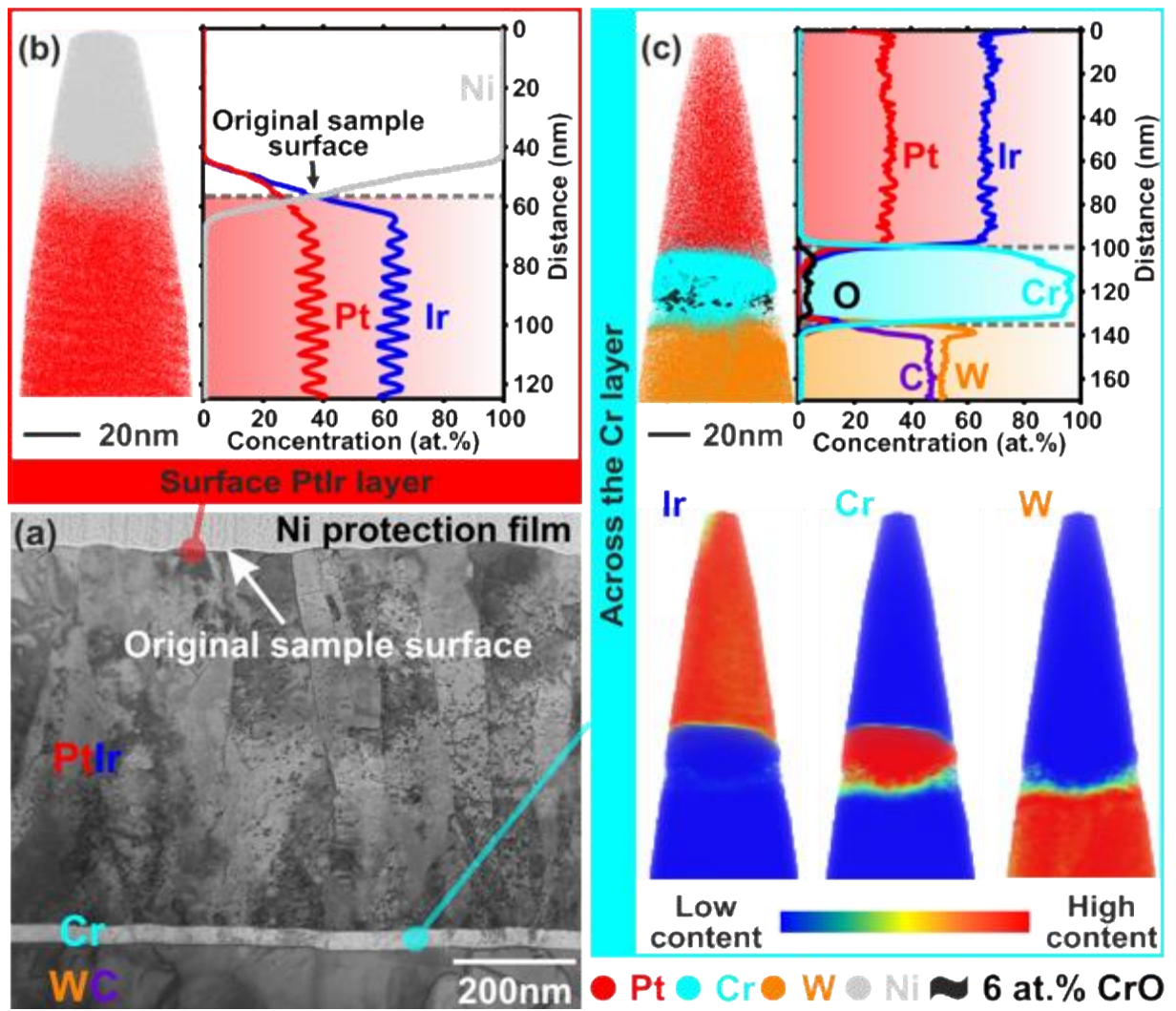

*Figure 6: Representative APT and STEM characterization results of an as-fabricated 600PtIr/20Cr sample: (a) cross-sectional bright field STEM image. (b) and (c) Cross-sectional atom map with the corresponding 1D composition-depth profile of the surface and the interfacial region respectively. In the Cr layer, some Cr oxides particles, marked using 6 at.% CrO iso-composition surface (black), are observed. To better illustrate the distribution of different elements within the sample, corresponding jet color composition maps of Ir, Cr, and W are also plotted in (c), where blue indicates the lowest content and red indicates the highest content. For clarity, in all atom maps, only Pt (red), Cr (cyan), W (orange) and Ni (light grey) atoms are shown.*

Figure 7 shows representative APT and STEM results obtained from the center region (position 3 marked in Figure 1 (a)) of a 600PtIr/20Cr sample after 120 PGM cycles in an industrial glass molding machine. Compared to the as-fabricated sample (Figure 6), we can find an additional ~30 nm thick Cr oxide scale on the surface, which is formed on the basis of the oxidation of Cr atoms that diffused through the grain boundaries of the PtIr layer to the surface [25]. Within the oxide scale, there is a small glass fragment with high Si content (Figure 7 (b)). The PtIr layer itself exhibits very good anti-sticking property against hot glass. However, as shown by previous studies [12,29,30], the newly formed oxide scale will enhance the chemical interaction between the coating and the glass blank, and lead to the adhesion of glass on the mold surface. Figure 7 (c) shows APT results from the PtIr layer, revealing that Sb atoms from the glass blank already penetrated into the PtIr coating through grain boundaries. Besides Cr and Sb, no other elements, such as O, were detected in a noticeable amount. Beyond main constituents of B270 i.e. $SiO_2$, $Na_2O$, $CaO$, $K_2O$, $BaO$, and $ZnO$, the glass also contains a very small amount of $Sb_2O_3$ (<1 wt.%) as refining agent to remove the unwanted bubbles from the glass melt during production [31].

In the interfacial region (Figure 7 (d)), bulk diffusion between the PtIr and Cr coating took place. As a result, the original Cr interlayer was replaced by an approx. 50 nm-thick interdiffusion zone (IDZ). On the bottom part of this IDZ, there is a small amount of W, which is a clear sign that the dissolution of the WC substrate already initiated. Although the sample history is different, the degradation phenomena observed here are very similar to that reported previously in a model system [25].

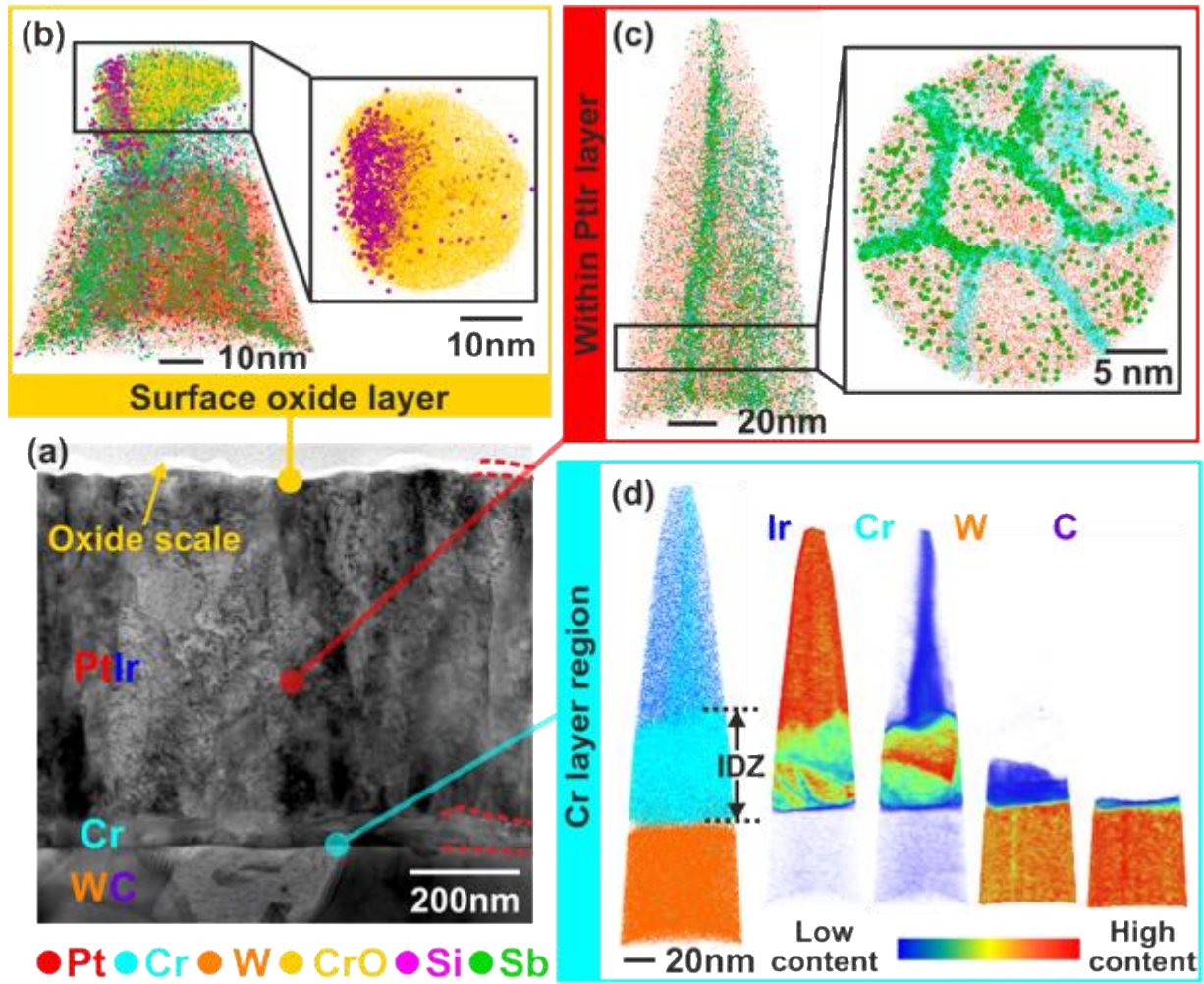

*Figure 7: Representative APT and STEM characterization results of the center region of a 600PtIr/20Cr sample after 120 PGM cycles in an industrial glass molding machine: (a) cross-sectional bright field STEM image. (b) Side view atom map of the surface region with a magnified top view of the region marked by the rectangle. (c) Side view atom map of the PtIr layer with a magnified top view of the region marked by the rectangle. (d) Cross-sectional view atom map of the interface region with the corresponding jet color composition maps of Ir, Cr, W and C, where blue indicates the lowest content and the red indicates highest content. The*

*original Cr layer was replaced by an interdiffusion zone (IDZ). In all atom maps, only Pt (red), Cr (cyan), W (orange), CrO (yellow), Si (magenta) and Sb (green) ions are shown for clarity.*

**3.2. Degradation during accelerated lifetime test**

Figure 8 shows the surface roughness Ra of a 600PtIr/20Cr sample during the test in the industrial glass molding machine as well as at the lifetime testing bench. The reported roughness for the industrial glass molding process represents the average of three roughness measurements performed at the positions marked 1–3 in Figure 1 (a). The roughness values for the lifetime testing bench process represent the average of nine roughness measurements done at the positions marked as 1-9 in Figure 1 (b). The error bars indicate the related standard deviations. The surface roughness of the sample increased slower during the industrial molding testing. After 120 molding cycles, the roughness value is only 7 nm, i.e. the mold is still qualified. While at the lifetime testing bench, only after 30 molding cycles, the surface roughness of the 600PtIr/20Cr sample is already 7 nm. After 90 molding cycles, it increased to 23.9 nm, far beyond the upper limit of the roughness required for production. This difference is supposed to result from the different atmospheres used in these two tests. The pressure in the industrial glass molding machine is approximately $3\times10^{-5}$ bar, which is much lower than that applied in the lifetime testing bench, i.e. $5\times10^{-3}$ bar.

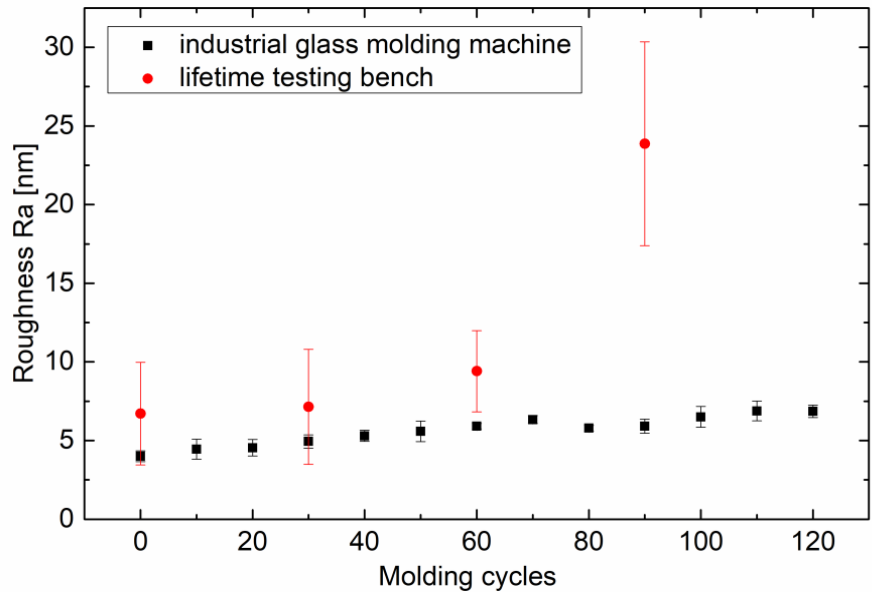

*Figure 8: Surface roughness of the 600PtIr/20Cr sample as a function of molding cycles during the test in the industrial glass molding machine as well as at the lifetime testing bench.*

Figure 9 shows light microscope micrographs and roughness measurements of a 600PtIr/20Cr sample before and after 30, 60 and 90 molding cycles at the lifetime testing bench. Each micrograph contains a very bright circle, which is the reflection of the microscope light, but not a surface feature of the sample. In the as-deposited state, the coating's surface was shining and defect free at the macroscopic scale. The surface roughness Ra was below 7 nm. After 30 molding cycles, the 600PtIr/20Cr sample shows a few bright points which are an indication of first macroscopic surface defects. The surface also starts to lose luster. After 60 molding cycles, the number and the size of surface defects increase. However, its surface roughness is only 9.4 nm, meaning the coating system is still qualified for operation. After 90 molding cycles, the 600PtIr/20Cr sample shows a dappled surface. Bright macroscopic areas alternate with areas providing the same dark color as deposited. The surface roughness value increases rapidly. It has more than doubled compared to after 60 molding cycles. The coating system is no more qualified for operation.

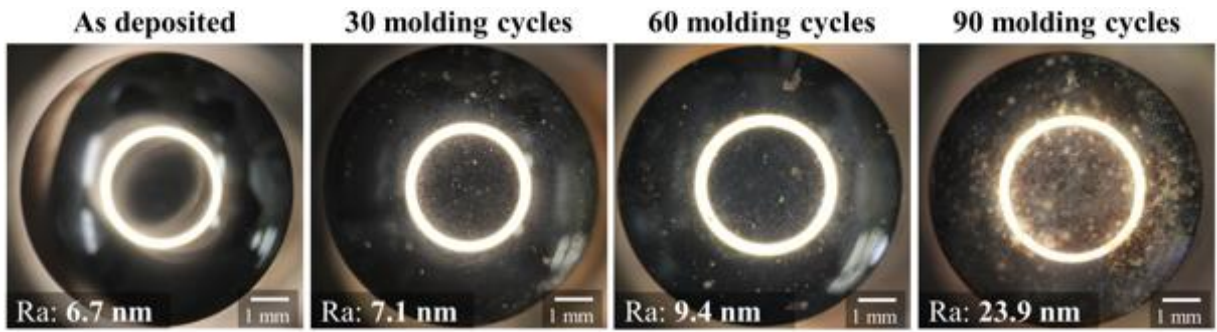

*Figure 9: Light microscope images and roughness of a 600PtIr/20Cr sample before and after 30, 60 and 90 molding cycles at the service lifetime testing bench.*

Similar to the 600PtIr/20Cr sample discussed above, all other studied coatings systems also show a progressive surface degradation with rising molding cycles. Figure 10 summarizes their roughness values after different molding cycles at the service lifetime testing bench. The surface roughness of all samples increases with rising number of molding cycles, as expected, but at different rates. The 300PtIr/20 sample degrades fastest, while the 600PtIr/20Cr sample degrades slowest. After 90 molding cycles, the roughness of the 300PtIr/20Cr sample is even higher than 550 nm, but that of the 600PtIr/20Cr is only 23.9 nm. Figure shows the light microscope images of the samples after 90 molding cycles. All of them show a brightening of the surface, which is an indication of severe surface defects. Similar to the 600PtIr/20Cr sample, the 600PtIr/5Cr sample shows a dappled surface, where the bright damaged areas predominate. The 600PtIr sample shows a severely damaged surface, evidenced by the brightening of almost the entire surface. The 300PtIr/20Cr sample shows the most massive surface damage compared to the other samples. The entire surface is damaged and roughed so much that the reflection of the microscope light is barely visible.

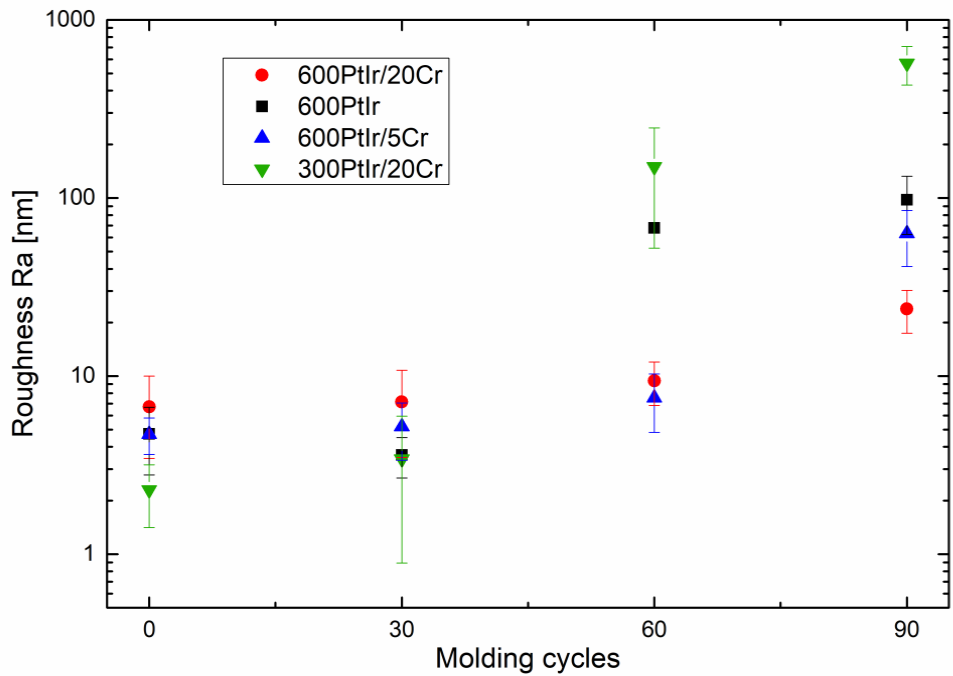

*Figure 10: Surface roughness of the 600PtIr (black squares), 600PtIr/5Cr (blue triangles), 600PtIr/20Cr (red dots), and 300PtIr/20Cr (green triangles) samples as a function of molding cycle during the test at the service lifetime testing bench.*

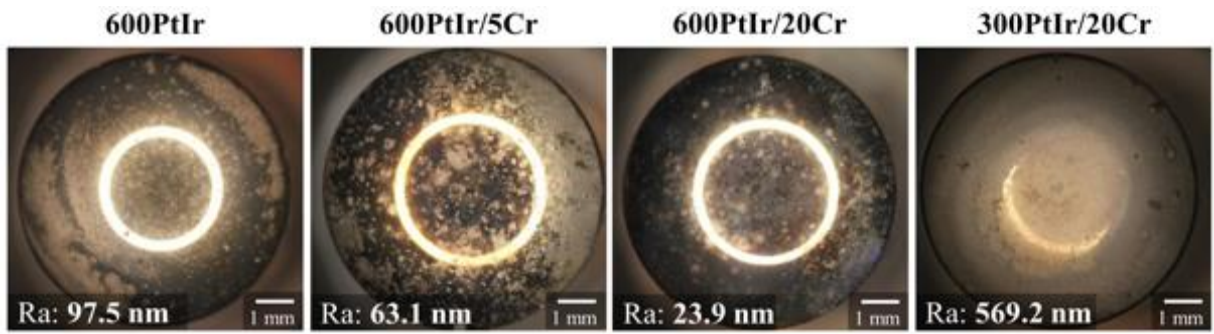

*Figure 11: Light microscope images and roughness of different samples after 90 molding cycles at the service lifetime testing bench.*

Figure 12 shows representative SEM and EDX results of (a) the 600PtIr, (b) 600PtIr/5Cr, (c) 600PtIr/20Cr and (d) 300PtIr/20Cr samples after 90 molding cycles at the service lifetime testing bench. On the surface of the 600PtIr sample, there are lots of small regions standing out. The EDX result (Figure 12 (a) spectrum 2) indicates that these area contains a large amount of

W from the substrate as well as K and Ca from the molded glass, which is a clear sign of glass adhesion. Outside these areas the protective coating is almost intact (spectrum 1).

SEM images of the 600PtIr/5Cr sample show with less and smaller local deposits over the surface than the 600PtIr sample. This explains the higher surface roughness of the 600PtIr sample compared to the 600PtIr/5Cr, shown in Figure 10. EDX analysis indicates again glass components, i.e. Ca, K, and Cl, in the deposits (Figure 12 (b) spectrum 1).

On the surface of the 300PtIr/20Cr sample, areas with an intact PtIr coating layer were detected (Figure 12 (d) spectrum 2). However, large areas of coating spallation were observed, which means that the substrate was directly exposed (spectrum 1). Elements from the glass, i.e. K and Zn, were also detected in that region.

The lower magnification SEM image of 600PtIr/20Cr coated surface in Figure 12 (c) indicates a mostly undamaged surface where only one large defect was identified. EDX analysis reveals that it contains several elements from the glass, i.e. P, Mg, Ca and a high amount of Si (spectrum 2). That high amount of Si was also noticed by APT characterization in a small glass fragment on the surface of the 600PtIr/20Cr sample after 120 molding cycles in an industrial glass molding machine, shown in Figure 7 (b). Nevertheless, the remaining surface has an intact PtIr coating layer (Figure 12 (c) spectrum 1). The 600PtIr/Cr20 sample show the least damage compared to the other PtIr coating systems studied, exhibiting the lowest surface roughness, as shown in Figure 10. Summarizing the accelerated lifetime tests, the 600PtIr/Cr20 coating system shows the best resistance to degradation compared to the other PtIr coating systems studied.

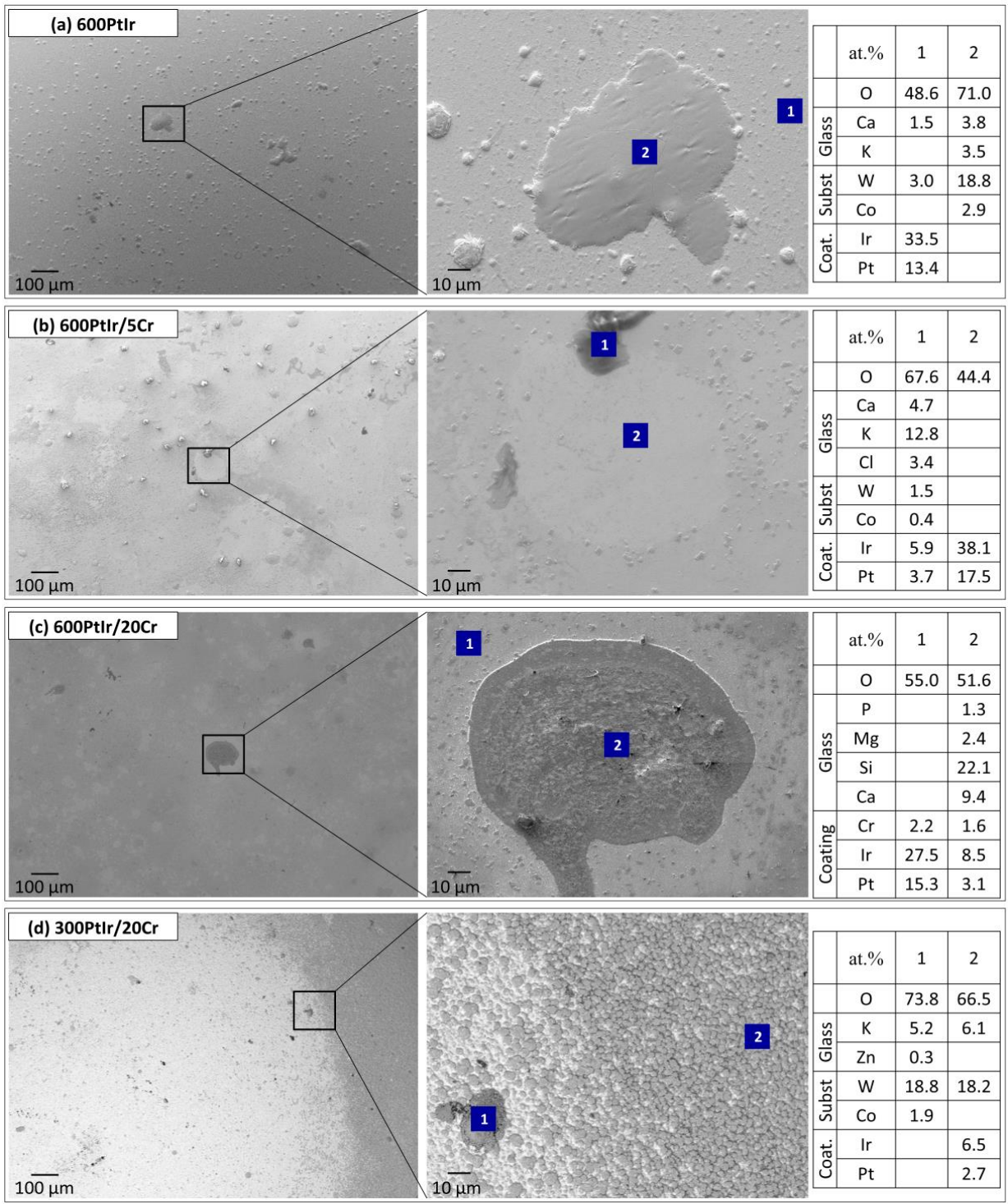

Figure 12: SEM images and EDX measurements of (a) the 600PtIr, (b) 600PtIr/5Cr, (c) 600PtIr/20Cr and (d) 300PtIr/20Cr samples after 90 molding cycles at the service lifetime testing bench.

## 4. Conclusion

The lifetime of 4 different PtIr protective coatings were evaluated by using an in-house designed lifetime testing bench and the detailed degradation mechanisms of the protective coating during the industrial PGM process were studied. The following conclusions can be drawn:

1) The in-house designed lifetime testing bench is economical and efficient in evaluating the lifetime of the coating.
2) The amount and size of the surface defects on the protective coatings increases with the increase of the molding cycles, as a result, their surface roughness increases.
3) Compared to the other coating systems studied in this work, the 600 nm PtIr protective coating with 20 nm Cr adhesion layer shows the best resistance to degradation.
4) Diffusion, surface oxidation, glass adhesion and coating flaking off restrict the lifetime of coated molding tools.
5) The grain boundaries of the PtIr layer work as fast diffusion channels for mass transport and accelerate the degradation process.

## Acknowledgements

The authors are grateful for the financial support from the InitialWear project, funded by the Fraunhofer-Gesellschaft (FhG) and the Max-Planck-Gesellschaft (MPG). The authors are thankful to Uwe Tezins & Andreas Sturm for their support of the APT & FIB facilities at MPIE.